\documentclass[aps,prb,twocolumn,showpacs]{revtex4}

\usepackage{graphicx}
\usepackage{amsmath}
\usepackage{amssymb}

\begin{document}

\title{Why there is a difference between optimal doping for maximal $T_{\rm c}$ and critical doping for highest
$\rho_{\rm s}$ in cuprate superconductors?}

\author{Zheyu Huang, Huaisong Zhao, and Shiping Feng$^{*}$}
\affiliation{Department of Physics, Beijing Normal University, Beijing 100875, China}

\begin{abstract}
A long-standing puzzle is why there is a difference between the optimal doping $\delta_{\rm optimal}\approx 0.15$ for
the maximal superconducting (SC) transition temperature $T_{\rm c}$ and the critical doping
$\delta_{\rm critical}\approx 0.19$ for the highest superfluid density $\rho_{\rm s}$ in cuprate superconductors? This
puzzle is calling for an explanation. Within the kinetic energy driven SC mechanism, it is shown that except the
quasiparticle coherence, $\rho_{\rm s}$ is dominated by the {\it bare} pair gap, while $T_{\rm c}$ is set by the
{\it effective} pair gap. By calculation of the ratio of the {\it effective} and the {\it bare} pair gaps, it is shown
that the coupling strength decreases with increasing doping. This doping dependence of the coupling strength induces a
shift from the critical doping for the maximal value of the {\it bare} pair gap parameter to the optimal doping for the
maximal value of the {\it effective} pair gap parameter, which leads to a difference between the optimal doping for the
maximal $T_{\rm c}$ and the critical doping for the highest
$\rho_{\rm s}$.
\end{abstract}


\maketitle

The parent compounds of cuprate superconductors are Mott insulators with an antiferromagnetic long-range order (AFLRO)
\cite{Kastner98}. However, this AFLRO is suppressed by doped charge carriers, then superconductivity arises from the
binding of charge carriers into Cooper pairs \cite{tsuei00}, thereby forming a superfluid with a superconducting (SC)
energy gap $\bar{\Delta}({\bf k})$ in the single-particle excitation spectrum. This energy gap is corresponding to the
energy for breaking a Cooper pair of the charge carriers and creating two excited states \cite{tsuei00}, while the
superfluid density $\rho_{\rm s}$ is proportional to the squared amplitude of the macroscopic wave function
\cite{schrieffer83}, and therefore describes the SC charge carriers. In this case, both $\bar{\Delta}({\bf k})$ and
$\rho_{\rm s}$ are thus two fundamental parameters whose variation as a function of doping and temperature provides
important information crucial to understanding the details of the SC state \cite{tsuei00,schrieffer83,bonn96}.

After intensive investigations over more than two decades, some essential features of the evolution of the SC state in
cuprate superconductors with doping have been experimentally established
\cite{bonn96,Hufner08,Presland91,ding01,niedermayer93,bernhard01,broun07}: where the measured energy gap parameter
$\bar{\Delta}$ and the SC transition temperature $T_{\rm c}$ show a domelike shape doping dependence, i.e., the maximal
$\bar{\Delta}$ and $T_{\rm c}$ occur around the {\it optimal doping} $\delta_{\rm optimal}\approx 0.15$, and then
decrease in both the underdoped and the overdoped regimes \cite{Hufner08,Presland91,ding01}. Moreover, the experimental
measurements \cite{niedermayer93,bernhard01,broun07} throughout the SC dome show that the superfluid density
$\rho_{\rm s}$ appears from the starting point of the SC dome, and then increases with increasing doping in the lower
doped regime. However, this $\rho_{\rm s}$ reaches its highest value around the {\it critical doping}
$\delta_{\rm critical}\approx 0.19$, and then decreases at the higher doped regime, eventually disappearing together
with $\bar{\Delta}$ at the end of the SC dome. In particular, it has been shown
\cite{bonn96,Hufner08,Presland91,ding01,niedermayer93,bernhard01,broun07} that the maximal $T_{\rm c}$ around the
{\it optimal doping} and the peak of $\rho_{\rm s}$ around the {\it critical doping} is a common feature of cuprate
superconductors. Since $\bar{\Delta}$ measures the strength of the binding of charge carriers into Cooper pairs
\cite{tsuei00}, while $\rho_{\rm s}$ is a measure of the phase stiffness \cite{bonn96}, therefore $\bar{\Delta}$ and
$\rho_{\rm s}$ separately describe the different aspects of the same SC charge carriers. In this case, a long-standing
puzzle is why there is a difference between the {\it optimal doping} for the maximal $T_{\rm c}$ and the {\it critical
doping} for the highest $\rho_{\rm s}$?

In this paper, we try to answer this question. Experimentally, the measured energy gap $\bar{\Delta}({\bf k})$ is an
{\it effective} energy gap \cite{Hufner08,Presland91,ding01}, which incorporates both the coupling strength and the
{\it bare} energy gap $\Delta({\bf k})$. Theoretically, the kinetic energy driven SC mechanism has been developed
\cite{feng0306}, where $T_{\rm c}$ is controlled by both the {\it effective} charge carrier pair gap and the
quasiparticle coherence. Within this kinetic energy driven SC mechanism, we calculate the doping dependence of the
coupling strength $V_{\rm eff}$, and the result shows that $V_{\rm eff}$ smoothly decreases upon increasing doping from
a strong-coupling case in the underdoped regime to a weak-coupling side in the overdoped regime. Our results also show
that the maximal value of the {\it bare} charge carrier pair gap parameter appears around the {\it critical doping}
$\delta_{\rm critical}\approx 0.195$, then as a natural consequence, the highest $\rho_{\rm s}$ occurs around this same
{\it critical doping}. However, the special doping dependence of $V_{\rm eff}$ shifts this {\it critical doping} for
the maximal value of the {\it bare} charge carrier pair gap parameter to the {\it optimal doping}
$\delta_{\rm optimal}\approx 0.15$ for the maximal value of the {\it effective} charge carrier pair gap parameter,
which leads to that $T_{\rm c}$ exhibits a maximum around the {\it optimal doping}.

Cuprate superconductors have a layered structure consisting of the two-dimensional CuO$_{2}$ planes separated by
insulating layers \cite{Kastner98}. The single common feature is the presence of the CuO$_{2}$ plane, and it seems
evident that the unusual behaviors of cuprate superconductors are dominated by this CuO$_{2}$ plane \cite{Kastner98}.
In this case, it has been argued that the essential physics of the doped CuO$_{2}$ plane is properly accounted by the
$t$-$J$ model on a square lattice \cite{anderson87}. However, for discussions of the difference between the
{\it optimal doping} for the maximal $T_{\rm c}$ and the {\it critical doping} for the highest $\rho_{\rm s}$, the
$t$-$J$ model can be extended by including the exponential Peierls factors as,
\begin{eqnarray}\label{tjham}
H&=&-t\sum_{l\hat{\eta}\sigma}P_{l\hat{\eta}}C^{\dagger}_{l\sigma}C_{l+\hat{\eta}\sigma}+t'\sum_{l\hat{\eta}'\sigma}
P_{l\hat{\eta}'}C^{\dagger}_{l\sigma}C_{l+\hat{\eta}'\sigma}\nonumber\\
&+&\mu\sum_{l\sigma} C^{\dagger}_{l\sigma}C_{l\sigma}+J\sum_{l\hat{\eta}}{\bf S}_{l}\cdot {\bf S}_{l+\hat{\eta}},
\end{eqnarray}
supplemented by an important on-site local constraint $\sum_{\sigma}C^{\dagger}_{l\sigma}C_{l\sigma}\leq 1$ to remove
the double occupancy, where the summation is over all sites $l$, and for each $l$, over its nearest-neighbors (NN)
$\hat{\eta}$ or the next nearest-neighbors (NNN) $\hat{\eta}'$, $C^{\dagger}_{l\sigma}$ and $C_{l\sigma}$ are electron
operators that respectively create and annihilate electrons with spin $\sigma$,
${\bf S}_{l}=(S^{\rm x}_{l},S^{\rm y}_{l},S^{\rm z}_{l})$ are spin operators, and $\mu$ is the chemical potential. The
exponential Peierls factors $P_{l\hat{\eta}}=e^{-i({e}/{\hbar}){\bf A}(l)\cdot\hat{\eta}}$ and
$P_{l\hat{\eta}'}=e^{-i({e}/{\hbar}){\bf A}(l)\cdot\hat{\eta}'}$ account for the coupling of electrons to an external
magnetic field in terms of the vector potential ${\bf A}(l)$ \cite{Abrikosov88}. To incorporate the electron single
occupancy local constraint in the $t$-$J$ model (\ref{tjham}), the charge-spin separation (CSS) fermion-spin theory
\cite{feng04,feng08} has been proposed, where a spin-up annihilation (spin-down annihilation) operator for the physical
electron is given by a composite operator as $C_{l\uparrow}=h^{\dagger}_{l\uparrow}S^{-}_{l}$
($C_{l\downarrow}=h^{\dagger}_{l\downarrow}S^{+}_{l}$), with the spinful fermion operator
$h_{l\sigma}=e^{-i\Phi_{l\sigma}}h_{l}$ that describes the charge degree of freedom of the electron together with some
effects of spin configuration rearrangements due to the presence of the doped hole itself (charge carrier), while the
spin operator $S_{i}$ represents the spin degree of freedom of the electron, then the electron single occupancy local
constraint is satisfied in analytical calculations. In this CSS fermion-spin representation, the $t$-$J$ model
(\ref{tjham}) can be rewritten as,
\begin{eqnarray}\label{cssham}
H&=&t\sum_{l\hat{\eta}}P_{l\hat{\eta}}(h^{\dagger}_{l+\hat{\eta}\uparrow}h_{l\uparrow}S^{+}_{l}S^{-}_{l+\hat{\eta}}
+h^{\dagger}_{l+\hat{\eta}\downarrow}h_{l\downarrow}S^{-}_{l}S^{+}_{l+\hat{\eta}})\nonumber\\
&-&t'\sum_{l\hat{\eta}'}P_{l\hat{\eta}'}(h^{\dagger}_{l+\hat{\eta}'\uparrow}h_{l\uparrow}S^{+}_{l}S^{-}_{l+\hat{\eta}'}
+h^{\dagger}_{l+\hat{\eta}'\downarrow}h_{l\downarrow}S^{-}_{l}S^{+}_{l+\hat{\eta}'})\nonumber\\
&-&\mu\sum_{l\sigma} h^{\dagger}_{l\sigma}h_{l\sigma}+J_{{\rm eff}}\sum_{l\hat{\eta}}{\bf S}_{l}\cdot
{\bf S}_{l+\hat{\eta}},
\end{eqnarray}
where $J_{{\rm eff}}=(1-\delta)^{2}J$, and
$\delta=\langle h^{\dagger}_{l\sigma}h_{l\sigma}\rangle=\langle h^{\dagger}_{l}h_{l}\rangle$ is the doping
concentration.

\begin{figure}[h!]
\includegraphics[scale=0.5]{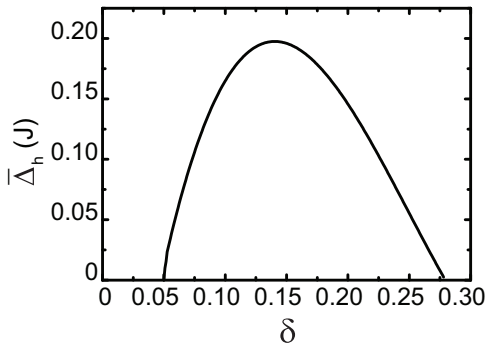}
\caption{The effective charge carrier pair gap parameter as a function of doping for temperature $T=0.002J$ with
parameters $t/J=2.5$ and $t'/t=0.3$.\label{fig1}}
\end{figure}

Since the experimental measurements \cite{shen93} have shown that in the real space the gap function and the pairing
force have a range of one lattice spacing, the {\it bare} energy gap parameter can be expressed as \cite{feng0306}
$\Delta=\langle C^{\dagger}_{l\uparrow}C^{\dagger}_{l+\hat{\eta}\downarrow}-C^{\dagger}_{l\downarrow}
C^{\dagger}_{l+\hat{\eta}\uparrow}\rangle=\langle h_{l\uparrow}h_{l+\hat{\eta}\downarrow}S^{+}_{l} S^{-}_{l+\hat{\eta}}
-h_{l\downarrow}h_{l+\hat{\eta}\uparrow}S^{-}_{l}S^{+}_{l+\hat{\eta}}\rangle$. In the doped regime without AFLRO, the
spin correlation functions
$\langle S^{+}_{l}S^{-}_{l+\hat{\eta}}\rangle=\langle S^{-}_{l}S^{+}_{l+\hat{\eta}}\rangle=\chi_{1}$, and then the
{\it bare} energy gap parameter can be rewritten as $\Delta=-\chi_{1}\Delta_{\rm h}$, with the {\it bare} charge
carrier pair gap parameter
$\Delta_{\rm h}=\langle h_{l+\hat{\eta}\downarrow}h_{l\uparrow}-h_{l+\hat{\eta}\uparrow}h_{l\downarrow}\rangle$,
which shows that the {\it bare} energy gap is closely related to the {\it bare} charge carrier pair gap, therefore the
essential physics in the SC state is dominated by the corresponding one in the charge carrier pairing state. For a
microscopic description of the SC state in cuprate superconductors, the kinetic energy driven SC mechanism has been
developed \cite{feng0306} based on the $t$-$J$ model (\ref{cssham}), where the charge carrier interaction directly from
the kinetic energy by exchanging spin excitations induces a d-wave charge carrier pairing state, and then their
condensation reveals the SC ground-state. Moreover, this SC state is controlled by both the {\it effective} energy gap
and the quasiparticle coherence. Within this kinetic energy driven SC mechanism, the full charge carrier Green's
function in the zero magnetic field case has been obtained explicitly in the Nambu representation as
\cite{feng08,guo07},
\begin{eqnarray}
\mathbb{G}({\bf k},i\omega_{n})=Z_{\rm hF}{i\omega_{n}\tau_{0}+\bar{\xi}_{\bf k}\tau_{3}-\bar{\Delta}_{\rm{hZ}}({\bf k})
\tau_{1}\over (i\omega_{n})^{2}-E_{{\rm h}{\bf k}}^{2}},
\label{holegreenfunction}
\end{eqnarray}
where $\tau_{0}$ is the unit matrix, $\tau_{1}$ and $\tau_{3}$ are Pauli matrices, the renormalized charge carrier
excitation spectrum $\bar{\xi}_{{\bf k}} =Z_{\rm hF}\xi_{\bf k}$, with the mean-field charge carrier excitation
spectrum $\xi_{{\bf k}}=Zt\chi_{1}\gamma_{{\bf k}}- Zt' \chi_{2}\gamma_{{\bf k}}'- \mu$, the spin correlation function
$\chi_{2}=\langle S_{l}^{+}S_{l+\hat{\eta}'}^{-}\rangle$,
$\gamma_{{\bf k}}=(1/Z)\sum_{\hat{\eta}}e^{i{\bf k}\cdot \hat{\eta}}$,
$\gamma_{{\bf k}}'= (1/Z)\sum_{\hat{\eta}'}e^{i{\bf k} \cdot\hat{\eta}'}$, $Z$ is the number of NN or NNN sites, the
renormalized charge carrier d-wave pair gap $\bar{\Delta}_{\rm hZ}({\bf k})=Z_{\rm hF} \bar{\Delta}_{\rm h}({\bf k})$,
and the charge carrier quasiparticle spectrum
$E_{{\rm{h}}{\bf k}}=\sqrt{\bar{\xi}^{2}_{{\bf k}}+ |\bar{\Delta}_{\rm hZ}({\bf k})|^{2}}$, where the {\it effective}
charge carrier d-wave pair gap $\bar{\Delta}_{\rm h}({\bf k})=\bar{\Delta}_{\rm h}({\rm cos} k_{x}-{\rm cos}k_{y})/2$,
and is closely related to the self-energy $\Sigma^{({\rm h})}_{2}({\bf k},\omega)$ in the particle-particle channel as
$\bar{\Delta}_{\rm h}({\bf k})=\Sigma^{({\rm h})}_{2}({\bf k},\omega)|_{\omega=0}$, while the quasiparticle coherent
weight $Z_{\rm hF}$ is directly associated with the self-energy $\Sigma^{({\rm h})}_{1}({\bf k},\omega)$ in the
particle-hole channel as $Z^{-1}_{\rm hF}=1-{\rm Re}\Sigma^{({\rm h})}_{\rm 1o}({\bf k},\omega=0)|_{{\bf k}=[\pi,0]}$,
with $\Sigma^{({\rm h})}_{\rm 1o}({\bf k},\omega)$ is the antisymmetric part of
$\Sigma^{({\rm h})}_{\rm 1}({\bf k},\omega)$, where the self-energies $\Sigma^{({\rm h})}_{1}({\bf k},\omega)$ and
$\Sigma^{({\rm h})}_{2}({\bf k},\omega)$ have been given in Refs. \onlinecite{feng08} and \onlinecite{guo07}. In this
case, the {\it effective} charge carrier pair gap parameter $\bar{\Delta}_{\rm h}$, $Z_{\rm hF}$, and all the other
order parameters have been determined by the self-consistent calculation \cite{feng08,guo07}. For a convenience in the
following discussions, the self-consistently calculated result \cite{guo07} of $\bar{\Delta}_{\rm h}$ versus doping for
temperature $T=0.002J$ with parameters $t/J=2.5$ and $t'/t=0.3$ is replotted in Fig. \ref{fig1}, where the maximal
$\bar{\Delta}_{\rm h}$ occurs around the {\it optimal doping} $\delta_{\rm optimal}\approx 0.15$, and then decreases
in both the underdoped and the overdoped regimes.

\begin{figure}[h!]
\includegraphics[scale=0.42]{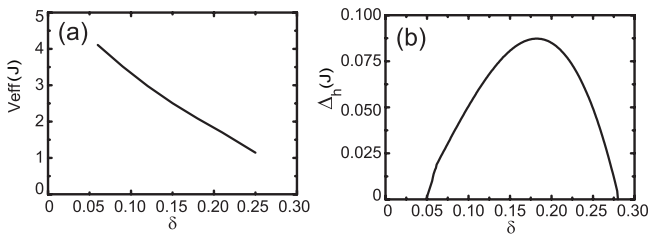}
\caption{(a) The coupling strength and (b) the {\it bare} charge carrier pair gap parameter as a function of doping
for temperature $T=0.002J$ with parameters $t/J=2.5$ and $t'/t=0.3$. \label{fig2}}
\end{figure}

\begin{figure}[h!]
\includegraphics[scale=0.4]{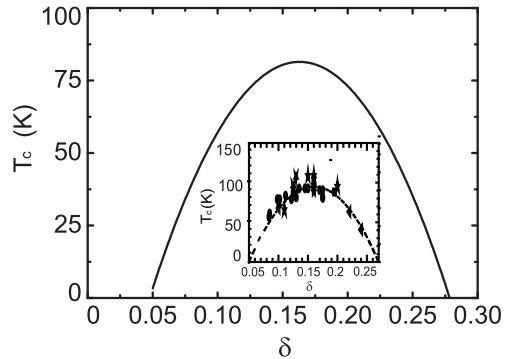}
\caption{The superconducting transition temperature as a function of doping with parameters $t/J=2.5$, $t'/t=0.3$, and
$J=1000$K. Inset: the corresponding experimental results of cuprate superconductors taken from Ref.
\onlinecite{Hufner08}. \label{fig3}}
\end{figure}

With the help of the Green's function (\ref{holegreenfunction}), the {\it bare} charge carrier pair gap parameter
$\Delta_{\rm h}$ can be evaluated explicitly as,
\begin{eqnarray}
\Delta_{\rm h}={1\over 2N}\sum_{{\bf k}}[{\rm cos} k_{x}-{\rm cos}k_{y}]^{2}{Z_{\rm hF}\bar{\Delta}_{\rm hZ}\over
E_{{\rm h}{\bf k}}}{\rm tanh}[{1\over 2}\beta E_{{\rm h}{\bf k}}].
\end{eqnarray}
Since the pairing force and this $\Delta_{\rm h}$ have been incorporated into the {\it effective} charge carrier pair
gap parameter $\bar{\Delta}_{\rm h}$ \cite{feng0306}, the strength $V_{\rm eff}$ of the attractive interaction mediated
by spin excitations in the kinetic energy driven SC mechanism can therefore be obtained in terms of the ratio of
$\bar{\Delta}_{\rm h}$ and $\Delta_{\rm h}$ as,
\begin{eqnarray}
V_{\rm eff}={\bar{\Delta}_{\rm h}\over\Delta_{\rm h}}.
\end{eqnarray}
Although both $\bar{\Delta}_{\rm h}$ and $\Delta_{\rm h}$ measure the strength of the binding of charge carriers into
charge carrier pairs, $\bar{\Delta}_{\rm h}$ is an experimentally measurable quantity, while $\Delta_{\rm h}$ is not.
In this case, we have calculated the doping dependence of $V_{\rm eff}$ and $\Delta_{\rm h}$, and the results of (a)
$V_{\rm eff}$ and (b) $\Delta_{\rm h}$ as a function of doping for $T=0.002J$ with $t/J=2.5$ and $t'/t=0.3$ are plotted
in Fig. \ref{fig2}, where the coupling strength $V_{\rm eff}$ smoothly decreases upon increasing doping from a
strong-coupling case in the underdoped regime to a weak-coupling side in the overdoped regime, which is consistent with
the experimental result of cuprate superconductors \cite{Kordyuk10}. However, $\Delta_{\rm h}$ increases with
increasing doping in the lower doped regime, and reaches a maximum around the {\it critical doping}
$\delta_{\rm critical}\approx 0.195$, then decreases with increasing doping in the higher doped regime. In comparison
with the corresponding result of $\bar{\Delta}_{\rm h}$ in Fig. \ref{fig1}, we therefore find that the special doping
dependence of the coupling strength $V_{\rm eff}$ in Fig. \ref{fig2}(a) induces an important shift from the
{\it critical doping} for the maximal $\Delta_{\rm h}$ to the {\it optimal doping} for the maximal
$\bar{\Delta}_{\rm h}$, then the doping dependence of $T_{\rm c}$ is determined by ${\bar{\Delta}_{\rm h}}$ (then both
$\Delta_{\rm h}$ and $V_{\rm eff}$) and the quasiparticle coherent weight $Z_{\rm hF}$ within the kinetic energy driven
SC mechanism \cite{feng0306}. To see this point clearly, $T_{\rm c}$ as a function of doping with $t/J=2.5$,
$t'/t=0.3$, and $J=1000$K is plotted in Fig. \ref{fig3} in comparison with the corresponding experimental results
\cite{Hufner08} of cuprate superconductors. It is shown that $T_{\rm c}$ increases with increasing doping in the
underdoped regime, and exhibits a maximum around the {\it optimal doping}, then decreases with increasing doping in the
overdoped regime, in good agreement with the experimental results of cuprate superconductors
\cite{Hufner08,Presland91,ding01}. In particular, $T_{\rm c}$ that is set by the {\it effective} pair gap and the
quasiparticle coherence has been observed experimentally in cuprate superconductors \cite{ding01}. We believe that
this property may be a common feature for all superconductors, since in spite of the electron-phonon SC mechanism,
$T_{\rm c}$ in the conventional superconductors is also determined by the {\it effective} pair gap and the
quasiparticle coherence \cite{eliashberg60}.

The essential physics of the domelike shape doping dependence of $T_{\rm c}$ in cuprate superconductors can be
attributed to a competition between the kinetic energy ($\delta t$) and magnetic energy ($J$) \cite{feng0306}. The
parent compounds of cuprate superconductors are the Mott insulators \cite{Kastner98}, when charge carriers are doped
into a Mott insulator, there is a gain in the kinetic energy per charge carrier proportional to $t$ due to hopping,
however, at the same time, the magnetic energy is decreased, costing an energy of approximately $J$ per site. As a
consequence, the strength of the spin excitation spectrum decreases with increasing doping, which leads to a decrease
of the coupling strength $V_{\rm eff}$ with increasing doping in the framework of the kinetic energy driven SC
mechanism. Moreover, in the underdoped regime, the coupling strength $V_{\rm eff}$ in Fig. \ref{fig2}(a) is very
strong, this implies that the most doped charge carriers can be bound into the charge carrier pairs, then the number
of the charge carrier pairs and $T_{\rm c}$ increase with increasing doping. However, in the overdoped regime, the
coupling strength $V_{\rm eff}$ is relatively weak. In this case, not all doped charge carriers can be bound to form
the charge carrier pairs by this weakly attractive interaction, and therefore the number of the charge carrier pairs
and $T_{\rm c}$ decrease with increasing doping. In particular, the {\it optimal doping} is a balance point, where
the number of the charge carrier pairs and the coupling strength $V_{\rm eff}$ are optimally matched. This is why the
$T_{\rm c}$ in cuprate superconductors exhibits a domelike shape doping dependence.

Now we turn to discuss the doping dependence of the superfluid density. The external magnetic field
${\bf B}=\rm{rot}{\bf A}$ applied to the system usually represents a large perturbation, but the induced field
generated by supercurrents cancels the external field over most of the volume of the sample. As a consequence, the net
field acts only very near the surface on a scale of the magnetic field penetration depth, and then it can be treated as
a weak perturbation on the system as a whole. In this case, the Meissner effect can be successfully studied within the
linear response approach \cite{fukuyama69}, where the response current density $J_{\mu}$ and the vector potential
$A_{\nu}$ are related by a nonlocal kernel of the response function $K_{\mu\nu}$ as,
\begin{equation}\label{linres}
J_\mu({\bf q},\omega)=-\sum\limits_{\nu=1,2,3}K_{\mu\nu}({\bf q},\omega)A_\nu({\bf q},\omega).
\end{equation}
This kernel of the response function in Eq. (\ref{linres}) can be separated into two parts as
$K_{\mu\nu}({\bf q},\omega)=K^{({\rm d})}_{\mu\nu}({\bf q},\omega)+K^{({\rm p})}_{\mu\nu}({\bf q},\omega)$, where
$K^{({\rm d})}_{\mu\nu}$ and $K^{({\rm p})}_{\mu\nu}$ are the corresponding diamagnetic and paramagnetic parts,
respectively, and are closely related to the current-current correlation function. The vector potential ${\bf A}$ has
been coupled to electrons, which are now represented by $C_{l\uparrow}= h^{\dagger}_{l\uparrow}S^{-}_{l}$ and
$C_{l\downarrow}= h^{\dagger}_{l\downarrow}S^{+}_{l}$ in the CSS fermion-spin representation. In this case, the
electron polarization operator is expressed as ${\bf P}=-e\sum\limits_{i\sigma}{\bf R}_{i}C^{\dagger}_{i\sigma}
C_{i\sigma}=e\sum\limits_{i\sigma}{\bf R}_{i}h^{\dagger}_{i} h_{i}$, and then the current operator ${\bf j}$ in the
presence of the vector potential $A_{\nu}$ is obtained by evaluating the time-derivative of this polarization operator.
According to this  current operator ${\bf j}$, the diamagnetic and paramagnetic parts of the response kernel have been
obtained in the static limit as \cite{feng10},
\begin{subequations}\label{kernel}
\begin{eqnarray}
K_{\mu\nu}^{(\rm d)}({\bf q},0)&=&-{4e^{2}\over\hbar^{2}}(\chi_{1}\phi_{1}t-2\chi_{2}\phi_{2}t')\delta_{\mu\nu}={1\over
\lambda^{2}_{L}}\delta_{\mu\nu},~~~~\label{diakernel} \\
K_{\mu\nu}^{(\rm p)}({\bf q},0) &=&{1\over N}\sum\limits_{{\bf k}}\gamma_{\mu}({\bf k}+{\bf q},{\bf k})\gamma^{*}_{\nu}
({\bf k}+{\bf q},{\bf k})[L_{1}({\bf k},{\bf q})\nonumber\\
&+&L_{2}({\bf k},{\bf q})]=K_{\mu\mu}^{(\rm p)}({\bf q},0)\delta_{\mu\nu},~~~~\label{parakernel}
\end{eqnarray}
\end{subequations}
where the charge carrier particle-hole parameters $\phi_{1}=\langle h^{\dagger}_{i\sigma}h_{i+\hat{\eta}\sigma}\rangle$
and $\phi_{2}=\langle h^{\dagger}_{i\sigma}h_{i+\hat{\eta}'\sigma}\rangle$,
$\lambda^{-2}_{L}=-4e^{2}(\chi_{1}\phi_{1}t- 2\chi_{2}\phi_{2}t')/\hbar^{2}$, while the functions
$L_{1}({\bf k},{\bf q},\omega)$ and $L_{2}({\bf k},{\bf q},\omega)$ have been given in Ref. \cite{feng10}. In
particular, we \cite{feng10} have shown that in the long wavelength limit, i.e.,
$|{\bf q}|\to 0$, $K_{yy}^{(\rm p)}({\bf q}\to 0,0)=0$ at $T=0$, reflecting that the long wavelength electromagnetic
response is determined by the diamagnetic part of the kernel only. However, at $T=T_{c}$,
$K_{yy}^{(\rm p)}({\bf q}\to 0,0)=-(1/ \lambda^{2}_{L})$, which exactly cancels the diamagnetic part of the response
kernel (\ref{diakernel}), and then the Meissner effect is obtained for all $T\leq T_{c}$. With the help of the response
kernel in Eq. (\ref{kernel}), the magnetic field penetration depth $\lambda(T)$ by taking into account the
two-dimensional geometry of cuprate superconductors within the specular reflection model has been evaluated as
\cite{feng10},
\begin{eqnarray}\label{lambda}
\lambda(T)={1\over B}\int\limits_{0}^{\infty}h_{z}(x)\,{\rm{d}}x={2\over\pi}\int\limits_{0}^{\infty}
{\rm{d}q_x\over\mu_{0}K_{yy}(q_{x},0,0)+q_{x}^{2}},
\end{eqnarray}
then the superfluid density $\rho_{\rm s}(T)$ is obtained as $\rho_{\rm s}(T)\equiv \lambda^{-2}(T)$.

\begin{figure}[h!]
\includegraphics[scale=0.26]{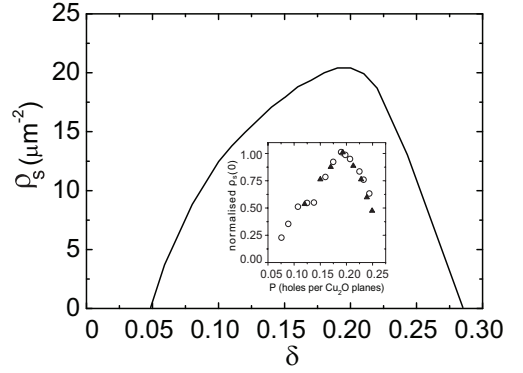}
\caption{The superfluid density as a function of doping for temperature $T=0.002J$ with parameters $t/J=2.5$,
$t'/t=0.3$, and $J=1000$K. Inset: the corresponding experimental results of cuprate superconductors taken from Ref.
\onlinecite{bernhard01}.
\label{fig4}}
\end{figure}

In this case, for the discussions of the difference between the {\it optimal doping} for the maximal $T_{\rm c}$ and
the {\it critical doping} for the highest $\rho_{\rm s}$, the result of $\rho_{\rm s}$ as a function of doping at
$T=0.002J$ with $t/J=2.5$, $t'/t=0.3$, and $J=1000$K is replotted in Fig. \ref{fig4} in comparison with the
corresponding experimental data \cite{bernhard01} of cuprate superconductors (inset). The result in Fig. \ref{fig4}
shows clearly that $\rho_{\rm s}$ increases with increasing doping in the lower doped regime, and reaches a maximum
around the {\it critical doping} $\delta_{\rm critical}\approx 0.195$, then decreases in the higher doped regime, in
good agreement with the experimental results of cuprate superconductors \cite{niedermayer93,bernhard01,broun07}. In
particular, this anticipated value of the {\it critical doping} $\delta_{\rm critical}\approx 0.195$ is very close to
the {\it critical doping} $\delta_{\rm critical}\approx 0.19$ obtained experimentally for different families of
cuprate superconductors \cite{niedermayer93,bernhard01,broun07}.

However, this {\it critical doping} $\delta_{\rm critical}\approx 0.195$ for the highest $\rho_{\rm s}$ is different
from that for $T_{\rm c}$ in Fig. \ref{fig3}, where the maximal $T_{\rm c}$ appears around the {\it optimal doping}
$\delta_{\rm optimal}\approx 0.15$. This difference can be understood within the present theoretical framework. Since
$\rho_{\rm s}$ is related to the current-current correlation function, the {\it bare} charge carrier pair gap
parameter $\Delta_{\rm h}$, the coupling strength $V_{\rm eff}$, and all the other order parameters are relevant, i.e.,
the variation of $\rho_{\rm s}$ with doping and temperature is coupled to the doping and temperature dependence of
$\Delta_{\rm h}$, $V_{\rm eff}$, and all the other order parameters \cite{feng10}. In particular, the doping-derivative
of $\rho_{\rm s}$ at the {\it critical doping} $\delta_{\rm critical}\approx 0.195$ is obtained as
$({\rm d}\rho_{\rm s}/{\rm d}\delta)|_{\delta=\delta_{\rm critical}}=0$. Since $\rho_{\rm s}\equiv\lambda^{-2}$,
$({\rm d}\rho_{\rm s}/{\rm d}\delta)|_{\delta=\delta_{\rm critical}}=0$ is equivalent to
$({\rm d}\lambda/{\rm d}\delta)|_{\delta=\delta_{\rm critical}}=0$. In this case,
$({\rm d}\lambda/{\rm d}\delta)|_{\delta=\delta_{\rm critical}}=0$ can be expressed in terms of Eq. (\ref{lambda}) as,
\begin{eqnarray}\label{dflambda}
\left [{{\rm d}\lambda\over {\rm d}\delta}\right ]_{\delta=\delta_{\rm critical}}&=&- {2\mu_{0}\over\pi}\int
\limits_{0}^{\infty}\rm{d}q_{x}\left [{1\over [\mu_{0} K_{yy}(q_{x},0,0)+q_{x}^{2}]^{2}}\right. \nonumber\\
&\times&\left. {{\rm d}K_{yy}(q_{x},0,0)\over {\rm d}\delta}\right]_{\delta=\delta_{\rm critical}}=0,
\end{eqnarray}
then it is straightforward to obtain from Eq. (\ref{kernel}) that when
$({\rm d}\rho_{\rm s}/{\rm d}\delta)|_{\delta=\delta_{\rm critical}}=0$,
$({\rm d}\Delta_{\rm h}/{\rm d}\delta)|_{\delta=\delta_{\rm critical}}=0$, which shows that the doping effects from
the coupling strength $V_{\rm eff}$ and all the other order parameters upon $\rho_{\rm s}$ are canceled each other,
then both the maximal $\Delta_{\rm h}$ and the highest $\rho_{\rm s}$ appear at the same {\it critical doping}.
Moreover, both $\rho_{\rm s}$ and $\Delta_{\rm h}$ are the {\it bare} quantities and separately describe the different
aspects of the same SC charge carriers. In this case, the domelike shape of the doping dependence of $\rho_{\rm s}$
with the highest value appeared around the {\it critical doping} is a natural consequence of the domelike shape of the
doping dependence of $\Delta_{\rm h}$ with the maximal value appeared around the same {\it critical doping} within the
kinetic energy driven SC mechanism. In other words, except the quasiparticle coherence, $\rho_{\rm s}$ is dominated by
the {\it bare} charge carrier pair gap parameter $\Delta_{\rm h}$, while $T_{\rm c}$ is set by the {\it effective}
charge carrier pair gap parameter ${\bar{\Delta}_{\rm h}}$, this is why there is a difference between the {\it optimal
doping} for the maximal $T_{\rm c}$ and the {\it critical doping} for the highest $\rho_{\rm s}$ in cuprate
superconductors. Finally, we have noted that $\rho_{\rm s}$ dominated by the {\it bare} energy gap parameter in cuprate
superconductors has been observed from the photoemission experiments \cite{he04}. Since in the SC state, the
photoemission peak intensity as a function of doping scales with $\rho_{\rm s}$, then a measurement of the coherent
component in the quasiparticle excitation has been suggested as an indirect measure of the {\it bare} energy gap
parameter in cuprate superconductors \cite{he04}.

In conclusion, within the framework of the kinetic energy driven SC mechanism, we have discussed the origin of the
difference between the {\it optimal doping} for the maximal $T_{\rm c}$ and the {\it critical doping} for the highest
$\rho_{\rm s}$ in cuprate superconductors. By calculation of the ratio of the {\it effective} and {\it bare} charge
carrier pair gap parameters, we have shown that the coupling strength decreases with increasing doping. This special
doping dependence of the coupling strength induces an important shift from the {\it critical doping}
$\delta_{\rm critical}\approx 0.195$ for the maximal value of the {\it bare} charge carrier pair gap parameter to the
{\it optimal doping} $\delta_{\rm optimal}\approx 0.15$ for the maximal value of the {\it effective} charge carrier
pair gap parameter, which leads to a difference between the {\it optimal doping} for the maximal $T_{\rm c}$ and the
{\it critical doping} for the highest $\rho_{\rm s}$ in cuprate superconductors.

\acknowledgments

This work was supported by the funds from the Ministry of Science and Technology of China under Grant Nos. 2011CB921700
and 2012CB821403, and the National Natural Science Foundation of China under Grant Nos. 11074023 and 11274044.

\end{document}